\documentclass[a4paper,accepted=2023-07-07]{quantumarticle}
\pdfoutput=1
\usepackage{amsmath}
\newcommand{\Haar}{\text{Haar}}
\newcommand{\Perm}{\text{Perm}}
\newcommand{\HOM}{\text{HOM}}
\DeclareMathOperator{\Tr}{Tr}
\DeclareUnicodeCharacter{2009}{\,} 
\DeclareUnicodeCharacter{2212}{-} 
\usepackage{graphicx}
\usepackage[separate-uncertainty=true]{siunitx}

\begin{document}

\title{20-Mode Universal Quantum Photonic Processor}

\author{Caterina Taballione}
\email{c.taballione@quixquantum.com}
\affiliation{%
 QuiX Quantum B.V., 7521 AN Enschede, The Netherlands
}%
\author{Malaquias Correa Anguita}
\affiliation{%
 MESA+ Institute for Nanotechnology, University of Twente,  7522 NB Enschede, The Netherlands
}%
\author{Michiel de Goede}
\affiliation{%
 QuiX Quantum B.V., 7521 AN Enschede, The Netherlands
}%
\author{Pim Venderbosch}
\affiliation{%
 QuiX Quantum B.V., 7521 AN Enschede, The Netherlands
}%
\author{Ben Kassenberg}
\affiliation{%
 QuiX Quantum B.V., 7521 AN Enschede, The Netherlands
}%
\author{Henk Snijders}
\affiliation{%
 QuiX Quantum B.V., 7521 AN Enschede, The Netherlands
}%
\author{Narasimhan Kannan}
\affiliation{%
 QuiX Quantum B.V., 7521 AN Enschede, The Netherlands
}%
\author{Ward L. Vleeshouwers}
\affiliation{%
 QuiX Quantum B.V., 7521 AN Enschede, The Netherlands
}%
\affiliation{%
 QuSoft, 1098 XG Amsterdam, The Netherlands
}%
\author{Devin Smith}
\affiliation{%
 QuiX Quantum B.V., 7521 AN Enschede, The Netherlands
}%
\author{Jörn P. Epping}
\affiliation{%
 QuiX Quantum B.V., 7521 AN Enschede, The Netherlands
}%
\author{Reinier van der Meer}
\affiliation{%
 MESA+ Institute for Nanotechnology, University of Twente, 7522 NB Enschede, The Netherlands
}%
\author{Pepijn W. H. Pinkse}
\affiliation{%
 MESA+ Institute for Nanotechnology, University of Twente, 7522 NB Enschede, The Netherlands
}%
\author{Hans van den Vlekkert}
\affiliation{%
 QuiX Quantum B.V., 7521 AN Enschede, The Netherlands
}%
\author{Jelmer J. Renema}
\affiliation{%
 QuiX Quantum B.V., 7521 AN Enschede, The Netherlands
}%
\affiliation{%
 MESA+ Institute for Nanotechnology, University of Twente, 7522 NB Enschede, The Netherlands
}%

\date{2023-07-07}

\begin{abstract}
Integrated photonics is an essential technology for optical quantum computing. Universal, phase-stable, reconfigurable multimode interferometers (quantum photonic processors) enable manipulation of photonic quantum states and are one of the main components of photonic quantum computers in various architectures. In this paper, we report the realization of the largest quantum photonic processor to date. The processor enables arbitrary unitary transformations on its 20 input modes with an amplitude fidelity of $F_{\Haar} = 97.4\%$ and $F_{\Perm} = 99.5\%$ for Haar-random and permutation matrices, respectively, an optical loss of \SI{2.9}{dB} averaged over all modes, and high-visibility quantum interference with $V_{\HOM}=98\%$. The processor is realized in $\mathrm{Si_3N_4}$ waveguides and is actively cooled by a Peltier element. 

\end{abstract}
\maketitle

\section{\label{sec:introduction}Introduction}
Photonics is one of the most attractive approaches to quantum computing, having gained momentum thanks to recent experimental results demonstrating a quantum advantage in photonics \cite{zhong_phase-programmable_2021, zhong_quantum_2020}. The strengths of photonic quantum computing platform are as follows: first, quantum states of light are characterized by inherently low decoherence due to their weak interaction with the surrounding environment; second, photonic quantum states maintain their coherence at room temperature; third, photonic quantum computing can exploit the high maturity of existing classical integrated photonics technologies. These factors together mean that integrated photonics represents a scalable approach to large-scale quantum computing. Finally, photons are the natural solution for  constructing quantum networks for either distributed quantum computing or for quantum communication \cite{kimble_quantum_2008, caleffi_quantum_2018, a_s_cacciapuoti_quantum_2020, illiano_quantum_2022}.

Quantum computational models based on photonics range from non-universal approaches, such as Boson Sampling \cite{aaronson_computational_2011}, which forms the basis of the recent quantum advantage experiments \cite{zhong_phase-programmable_2021, zhong_quantum_2020}, to universal ones based on a variety of different encodings  \cite{kok_linear_2007,takeda_toward_2019, andersen_hybrid_2015, larsen_deterministic_2021,slussarenko_photonic_2019,flamini_photonic_2018}. Applications of non-universal photonic quantum computing have been proposed, such as quantum chemistry \cite{sparrow_simulating_2018, banchi_molecular_2020} and graph properties \cite{bromley_applications_2019}. 

Linear optics is at the core of photonic quantum computing, as it is the natural means to generate entanglement between photons. Despite the fact that photons are non-interacting particles, entanglement can be formed via quantum interference in a linear optical system, with the archetypal example being the Hong-Ou-Mandel effect \cite{hong_measurement_1987}. One can place several requirements on a linear optical system for quantum interference. First, the system must be programmable and universal, in the sense that arbitrary optical transformations can be set by the user, with high fidelity. Second, low losses are also a prerequisite for photonic quantum computing as otherwise the information carried by quantum light is lost. Third, the linear optical system must be large in scale, to increase the complexity of the calculation that can be performed.

Integrated photonics has become essential for photonic quantum computing \cite{wang_integrated_2020} as it represents a scalable, mature and commercially available tool to realize large-scale, inherently phase-stable and fully-reconfigurable linear optical interferometers, called a Quantum Photonic Processor (QPP) in this context.
A quantum photonic processor \cite{sparrow_simulating_2018,carolan_universal_2015,harris_quantum_2017,taballione_88_2019,taballione_universal_2021} is one of the essential components of photonic quantum computing and is the main player in applications such as quantum neural networks \cite{steinbrecher_quantum_2019}, quantum metrology \cite{matthews_towards_2016}, PUFs \cite{smith_reconfigurable_2020},  witnesses of bosonic interference \cite{van_der_meer_experimental_2021,brod_witnessing_2019} and for benchmarking multi-photon light sources \cite{tiedau_statistical_2021,pont_quantifying_2022}.
Of the available integrated platforms, stoichiometric silicon nitride $\mathrm{Si_3N_4}$ is the most promising platform for photonic quantum computing. It provides an optimal combination of low loss and high optical mode confinement \cite{c_g_h_roeloffzen_low-loss_2018}, enabling the scaling up of low-loss fully-reconfigurable linear optical interferometers for quantum computing and information processing.

In this paper, we present the largest universal quantum photonic processor to date, with 20 input/output modes. Throughout this paper, ‘modes’ refers to the input/output waveguides of the processor, supporting only the fundamental optical mode. We note that each waveguide supports optical modes at multiple frequencies.

Our processor is based on stoichiometric silicon nitride $\mathrm{Si_3N_4}$ waveguides using TripleX technology \cite{triplex} and is backed with a water-cooled Peltier element to maintain a constant temperature. The waveguides are realized with an asymmetric double-stripe cross-section and have a minimum bending radius of 100 $\mu$m. The device has losses of \SI{2.9}{dB} averaged over all modes and enables arbitrary linear optical transformations, making it compatible with all linear optical models of quantum computation. We test the reconfigurability of the processor over more than 1000 unitary transformations, and show a high degree of control of the linear optical interferometer, as well as its universality. We further validate the processor with 190 quantum interference experiments confirming that the processor preserves the properties of quantum light at each of its components.

\section{\label{sec:chip} The Quantum Photonic Processor}
\begin{figure}[tbh]
  \includegraphics[scale=0.94]{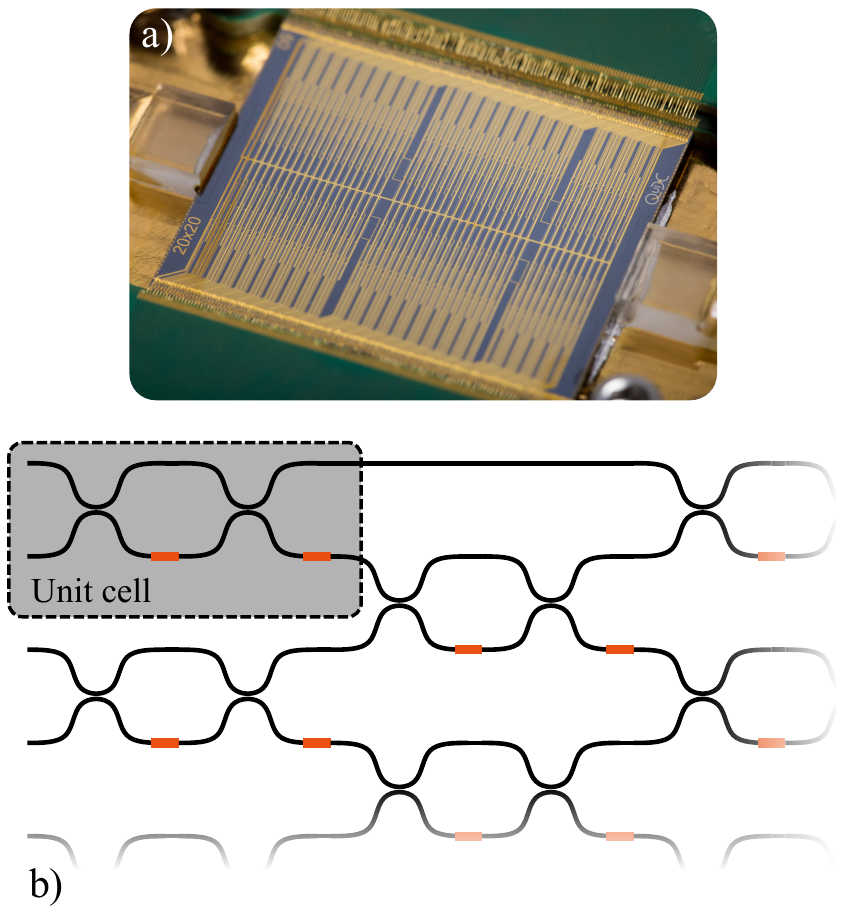}
  \caption{\textbf{a)} Photograph of the 20-mode processor chip ($\SI{22}{mm} \times \SI{30}{mm}$). The chip is optically packaged to an input/output fiber array and it is wire-bonded to the control PCB enabling the addressing of each individual tunable element. \textbf{b)} Functional design of the processor displaying the outline of the thermo-optic tunable elements (in red) and ideal layout of the waveguide paths (in black). The mesh is built by repeating the unit cell, comprising a tunable beam splitter (TBS) and external phase shifter (PS), $\frac{1}{2}N(N-1)$ times where $N=20$ for a total of 190 unit cells.}
  \label{fig:chip}
\end{figure}

Our 20-mode quantum photonic processor 
consists of three parts: the $\mathrm{Si_3N_4}$ photonic chip, the peripheral system which includes the control electronics, and dedicated control software \cite{taballione_universal_2021}.
The $\mathrm{Si_3N_4}$ photonic chip (Fig.\ref{fig:chip}a) contains a total of 380 thermo-optic tunable elements, arranged in a universal square interferometer (Fig. \ref{fig:chip}b) where the unit cell comprises a tunable beam splitter (TBS) followed by a phase shifter (PS) \cite{clements_optimal_2016}. Propagation losses as low as \SI{0.07}{dB/cm} at \SI{1562}{nm} are measured across the entire photonic chip, obtained by a higher-temperature annealing process compared to our previous chip \cite{taballione_universal_2021}.
The peripheral system comprises the control electronics and an active cooling module. The thermo-optic tunable elements can be switched at $\si{kHz}$ rate \cite{c_g_h_roeloffzen_low-loss_2018}, setting the limit for the switching speed between different configurations of the processor. To precisely control the temperature of the photonic chip, it is actively cooled. The cooling module consists of a Peltier element attached to a water cooling module providing a maximum heat reduction rate of \SI{200}{W}.

The dedicated control software performs both the decomposition of any unitary matrix transformation into the phase settings of each unit cell and their assignment to the corresponding tunable elements. The control software takes also into account the imperfections of the processor, such as crosstalk of individual tunable elements and compensates for those. 

\section{\label{sec:setup} Experimental Results}
In order to demonstrate the suitability of our processor for photonic quantum computing, we demonstrate full reconfigurability and control, as well as preservation of quantum interference across the whole 20-mode linear optical interferometer.  

\subsection{\label{sec:classical_results}Classical Results}
\begin{figure}
  \includegraphics[scale=0.94]{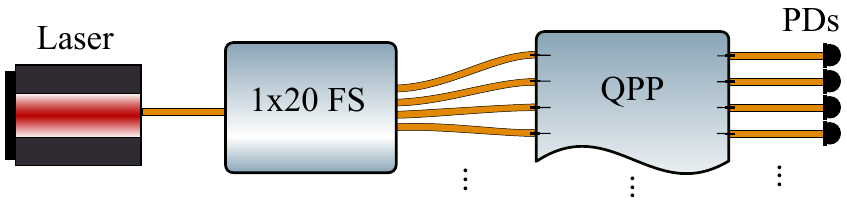}
  \caption{Sketch of the experimental setup for classical characterization of the processor. Laser: Santec TSL-570, FS: fiber switch Fibermart $1\times32$ PM, PDs: Thorlabs FGA01FC}
  \label{fig:setupclassical}
\end{figure}

The full control of the processor is demonstrated by characterizing the phase-voltage relationship of each thermo-optic actuator. This step is performed by injecting coherent light into the 20-mode processor via a $1\times20$ PM fiber switch (Fig.\,\ref{fig:setupclassical}). 

Each tunable element is driven to discrete voltage values and its output power is recorded at a photodiode array (PDs) and processed by the control software to obtain the true phase response of each tunable element.  The characterization of the 380 tunable elements shows that they all have a phase tuning range exceeding $2\pi$, allowing for full control of the unitary transformation implemented within the interferometer. The insertion loss of the photonic processor is measured to be  \SI{2.9(2)}{dB}: this is the overall loss experienced by light going in and out of the processor, from input to output fiber, through all the tunable MZIs and phase shifters. The measured insertion loss corresponds to coupling and propagation loss of, respectively, \SI{0.9}{dB/facet} and \SI{0.07}{dB/cm}.

\begin{figure*}
\includegraphics[scale=0.94]{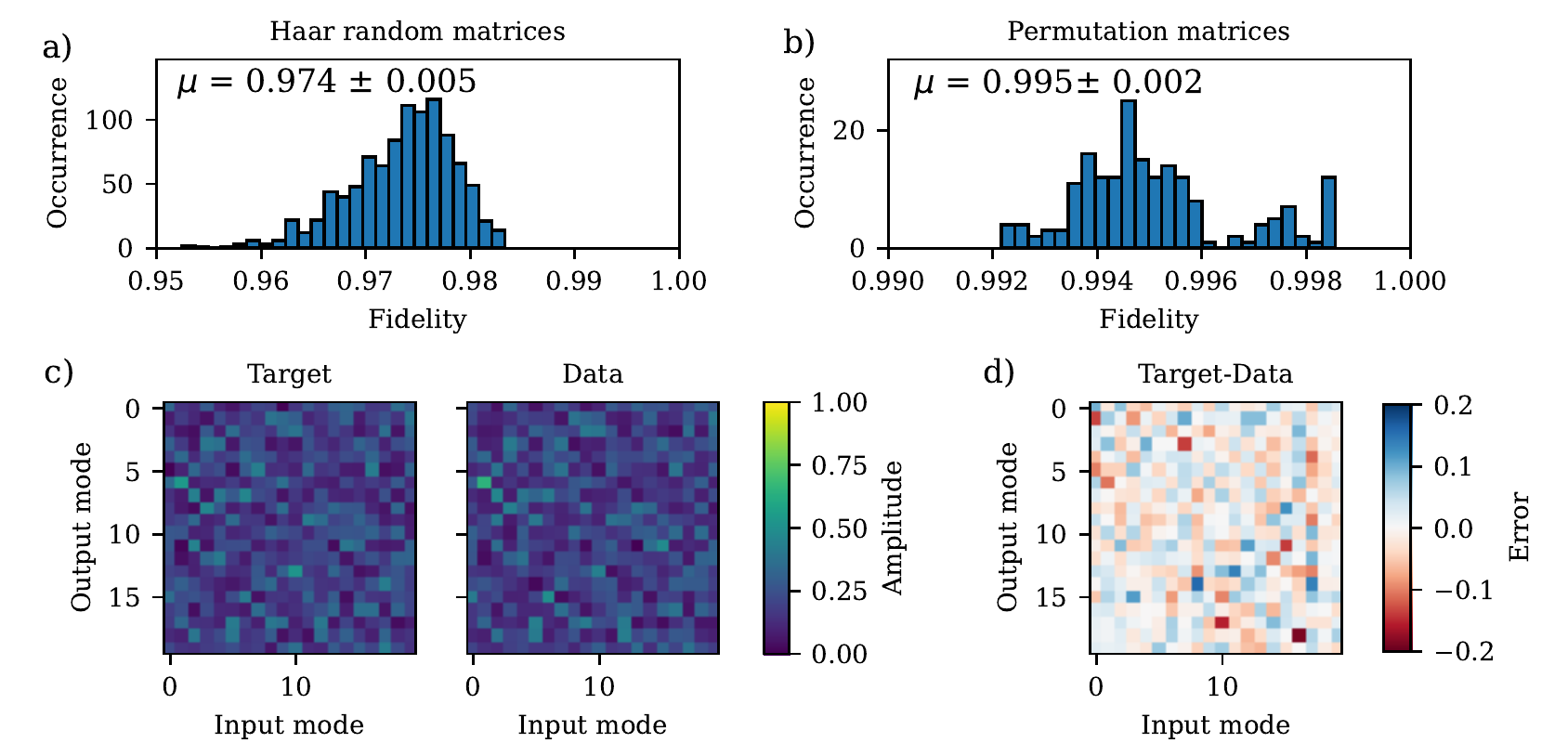}
  \caption{Summary of the classical characterization. \textbf{a)} and \textbf{b)} Distribution of the measured amplitude fidelities for, respectively, 1000 Haar-random matrices  of average fidelity 
  \SI{97.4(5)}{\%} and 190 Permutation matrices with average fidelity \SI{99.5(2)}{\%}. \textbf{c)} Comparison between the amplitude distribution of an ideal Haar-random unitary matrix (target) with its experimental implementation (data). \textbf{d)} Difference between the target and the measured (data) Haar random matrix.}
  \label{fig:haar-comparison}
\end{figure*}

Finally, we verify the reconfigurability and control of the processor by generating and implementing 190 permutation and 1000 Haar-random matrices on the device. Generating Haar-random unitary matrices is done via the method proposed in \cite{mezzadri}. For each input mode of the permutation and Haar-random matrices, the output intensity distribution is measured. From this distribution, a fidelity measure $F = 1/N\, \Tr(|U^+|\cdot|U_{\rm exp}|)$ of the unitary optical transformations on the input light is determined, where $|U|$ refers to the component-wise absolute value and where $N = 20$ is the number of modes. This fidelity measure is typically referred to as the amplitude fidelity. We obtain such fidelities as high as  $F = $ \SI{99.5(2)}{\percent} and  $F = $ \SI{97.4(5)}{\%}, for the permutations and the Haar-random transformations, respectively  (see also Fig.\,\ref{fig:haar-comparison}a and b).

As an example, Fig.\,\ref{fig:haar-comparison}c shows a comparison between one of the target Haar-random matrices (left), and the measured results from the implementation of that matrix on the QPP (right). The strong resemblance between the two plots can be clearly seen, as confirmed by the matrix plot in Fig.\,\ref{fig:haar-comparison}d 
showing that the error on the implemented amplitude matrix elements falls within 0.2.

We note that the processor's performance remains stable for at least 6 months, i.e., there is no need of re-calibrating the device for at least half a year. This is due to the stability of the on-chip structure and control electronics. 

\subsection{\label{sec:quantum_results}Quantum Results}
Measuring the visibility of HOM interference \cite{hong_measurement_1987} at every location on the processor provides quantum validation of the device. In order to do so, we use a photon-pair source based on degenerate Type-II spontaneous parametric down-conversion (SPDC) in periodically poled potassium titanyl phosphate (PPKTP). Pumped at \SI{775}{nm} with picosecond pulses, pairs of photons are generated with low probability on each pulse. 
The generated photons are highly indistinguishable and frequency-unentangled, with a mutual Schmidt number of about 1.1. For this experiment, the photons are filtered to $\Delta \lambda \approx \SI{12}{nm}$ using bandpass filters, to ensure maximum purity of the two-photon state. The photons are collected by optical fibers. Partial temporal distinguishability between the photons can be continuously tuned by varying the relative path length of the photons using a motorized linear displacement stage. 

\begin{figure}[tbh]
  \includegraphics{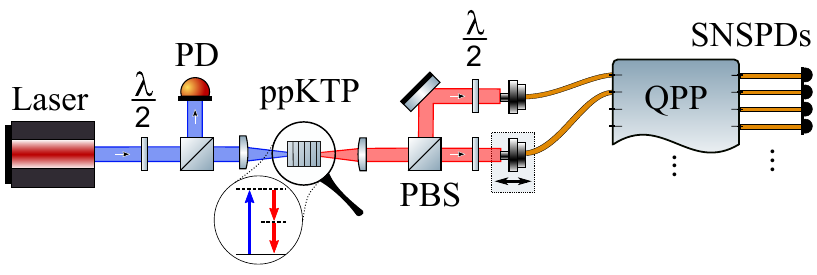}
  \caption{Sketch of the experimental setup for quantum characterization of the processor. A Ti:Sapphire laser is used to pump a ppKTP crystal, producing a pair of degenerate photons. The photons are injected into the processor where HOM interefence is tested. Each output mode is connected to an SNSPD.}
  \label{fig:Setup_HOM}
\end{figure}

The photons are then injected into pairs of input modes, as in Fig.\,\ref{fig:Setup_HOM}, and are routed to interfere at each on-chip TBS, set to a 50:50 splitting ratio.
The output modes are connected to superconducting nanowire single-photon detectors (SNSPDs), whose coincidence rate is measured using a standard time-tagger.

\begin{figure*}[tb]
  \includegraphics[scale=0.94]{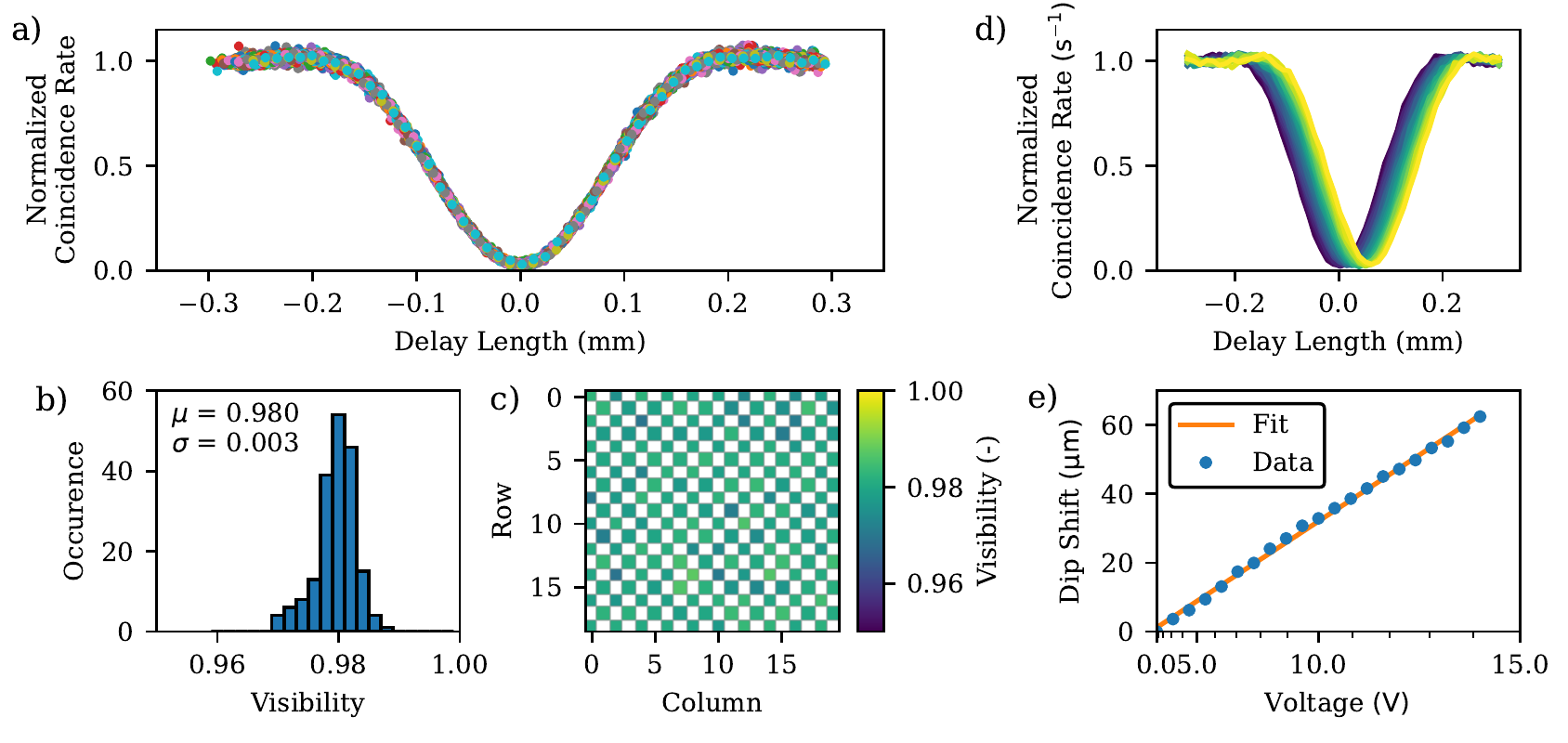}
  \caption{Results of the HOM interference measurements for all TBS on the processor. \textbf{a)} The normalized coincidence rates of all TBS in the processor; 190 separate interferograms are shown. A Gaussian fit was used to determine the visibility of the HOM interference. \textbf{b)} Histogram of the HOM visibility of all TBS in the processor. \textbf{c)} HOM visibilities of all TBS for each processor row and column. The checkerboard pattern reflects the alternating pairwise coupling of neighbouring channels as laid out in Fig.\,\ref{fig:chip}b. \textbf{d)} Normalized coincidence rates for various delays induced by increasing voltages across the main diagonal. \textbf{e)} Total path length tunability of the longest diagonal path across the matrix.}
  \label{fig:visibilities}
\end{figure*}

In order to route the two photons to every on-chip TBS, and to the connected outputs, the entire interferometer is used. TBSs are set to full reflection or transmission to create optical paths, which contain no intersections other than the TBS of interest.
The normalized HOM dips at all 190 TBS are reported in Fig.\ref{fig:visibilities}a: the clear overlap of all plots shows that the processor preserves the indistinguishability of the input single photons, as can be also seen in the low spread of the visibility histogram in Fig.\,\ref{fig:visibilities}b. The spatial distribution of the HOM visibilities over the rows and columns of the processor, as shown in Fig.\ref{fig:visibilities}c, is quite random, confirming that there are no  systematic errors within the processor.  Furthermore, the visibility of the HOM interference appears to be limited by the quality of the source used for the characterization.

To preserve the quantum interference, it is important that the overlap between photons remains as high as possible. Any path length difference, either geometrical or optical, will lower this overlap. In multi-photon experiments, single photons can interfere not only on a specific TBS but also across the entire processor via different paths that might be quite different from each other. It is thus important to quantify the effect of on-chip path length differences. %

To assess this, we measure the path length differences that our processor induces over a long on-chip path, given by the shift of the HOM interference dip obtained by injecting two single photons in the top two inputs and measuring their coincidences at the bottom two outputs, i.e., over the main diagonal of the processor. The induced path length difference is varied by sweeping the voltage of every phase shifter on one of the arms of the main diagonal. This induced path length difference could be observed as a shifting of the HOM dip positions, see Fig.\ref{fig:visibilities}d and Fig.\ref{fig:visibilities}e. A shift of more than \SI{60}{\micro\meter} is measured, or a maximal phase shift of at least $3\pi$ per heater. Depending on the coherence length of the photons, this temporal delay can induce a strong effect on the indistinguishability of the photons.

\section{\label{sec:discussion_conclusions}Discussion and Conclusions}

The current quantum photonic processor distinguishes itself from its predecessor \cite{taballione_universal_2021} in several ways (Table \ref{table:data}). 
\begin{table}
\centering
\footnotesize
\begin{tabular}{l c c c c} 
  & $N$ & IL (dB) & CL (dB/facet) & PL (dB/cm) \\  
\hline
12-mode & 132 & 5 & 2.1 & 0.1 \\
20-mode & 380 & 2.9 & 0.9 & 0.07 \\
\end{tabular}

\caption{Comparison between the 12-mode \cite{taballione_universal_2021} and 20-mode processor. $N$ = number of thermo-optic phase shifters, IL = Insertion Loss, CL = Coupling Loss, PL = Propagation Loss}
\label{table:data}
\end{table}
On the hardware level, we increased the number of optical input and output modes from 12 to 20. Despite its greater size and the increased optical path lengths per mode, the insertion loss is significantly reduced to an average of \SI{2.9(2)}{dB}. This reduction stems mostly from improved fiber-to-chip coupling with an average coupling loss of \SI{0.9}{dB/facet}.

A quantum photonic processor is at the core of a photonic quantum computer. Crucial properties of a QPP include its full programmability and universality, which we demonstrate in this work. Other critical properties for optimal quantum processing are low losses and large scale, i.e., large number of unit cells. These two features are not always compatible with each other as usually high-index-contrast high-confinement material platforms that enable the densest optical circuits (largest number of unit cells on a wafer), such as SOI or InP, suffer from greater losses (both propagation and coupling) than other platforms, such as silicon nitride (SiN) or silica. On one hand, this is due to the strong interaction of the optical mode with the sidewall surface roughness and, on the other hand, to the large optical mode mismatch at the facets of the chip. In the next paragraphs we discuss and show the SiN technology as the best platform for realizing universal quantum photonic processors. SiN is in fact capable of satisfying the requirement of large-number of modes while, at the same time, providing ultra-low losses. 

\begin{figure}
  \includegraphics[scale=0.94]{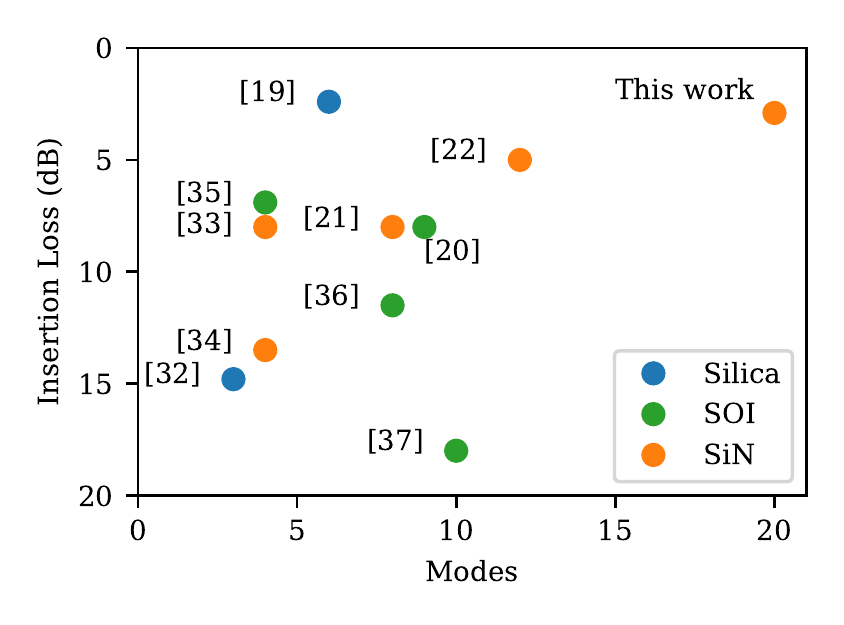}
  \caption{Overview of universal photonic processors performances as reported in the literature. We plot the insertion loss of each processor versus the number of optical modes of the largest universal transformation as in \cite{clements_optimal_2016} that can be implemented within each processor. Insertion loss includes the on-chip propagation loss and twice the fiber-to-chip coupling loss.}
  \label{fig:ILvsModes}
\end{figure}

If we locate our work in the context of universal photonic processors, i.e., fully-reconfigurable and all-to-all connectivity, it can be clearly seen that our systems provide the highest number of input/output modes for the lowest insertion losses (Fig.\ref{fig:ILvsModes}). 
Insertion losses, in this case, include both coupling and on-chip losses. This confirms the prominence of silicon nitride technology as mature platform for realizing universal quantum photonic processors. The data points in Fig.\ref{fig:ILvsModes} come from \cite{carolan_universal_2015, mennea_modular_2018} for silica, \cite{taballione_88_2019, taballione_universal_2021, arrazola_quantum_2021,de_marinis_photonic_2021} for SiN and \cite{harris_quantum_2017, ribeiro_demonstration_2016, zhang_optical_2021, tang_ten-port_2021} for silicon-on-Insulator. Other impressive works of large-scale photonic integrated circuits can be found in the literature, however they are either non-universal or have unknown performance in terms of losses \cite{shadbolt_generating_2012,santagati_silicon_2017, bell_testing_2019,seok_wafer-scale_2019,feldmann_parallel_2021,hoch_boson_2021,qiang_implementing_2021,wang_ultra-wide_2021,suzuki_low-loss_2020,ding_reconfigurable_2016,dong_high-speed_2022,crespi_integrated_2013,bentivegna_bayesian_2014, harris_quantum_2017,qiang_large-scale_2018,spring_boson_2013,tillmann_experimental_2013, zhou_self-configuring_2020, vigliar_error-protected_2021}.

\begin{figure}
  \includegraphics[scale=0.94]{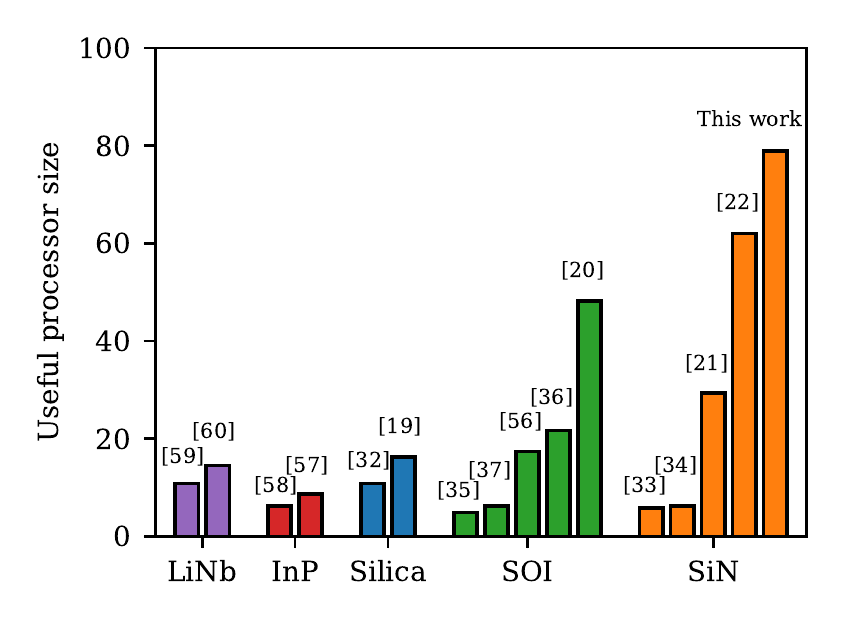}
  \caption{Overview of the useful processor size for various material platforms. We consider the loss per unit cell of each cited paper and plot the number of unit cells that can be sequentially concatenated before reducing the transmission to $\mathrm{e^{-1}}$.}
  \label{fig:platformcomparison}
\end{figure}

Scalability is another crucial property to consider when dealing with large-scale universal photonic processors.
On-chip propagation losses are to be considered the main limiting factor. In particular, if propagation losses are too high, no useful universal processors can be realized.
In order to compare the scalability of various material platforms, we report in Fig.\ref{fig:platformcomparison} the achievable processor size before reaching an arbitrary loss value, i.e., the useful processor size.
We take the loss per unit cell of various processors as found in the literature and calculate the processor size at which the transmission is reduced to an arbitrary threshold, in this case $\mathrm{e^{-1}}$. The reader should note that computing or information processing protocols have different transmission thresholds. The reader should also note that the useful processor size in Fig.\ref{fig:platformcomparison} is based on the current state of the art and should therefore not be considered the absolute upper limit: as the loss per unit cell decreases, the useful processor size increases. The data points in Fig.\ref{fig:platformcomparison} are the same as in Fig.\ref{fig:ILvsModes} with the addition of \cite{annoni_unscrambling_2017} for SOI. For completeness of the overview, we include data points for active materials such as indium phosphate \cite{q_cheng_scalable_2013,thiessen_30_2019} and lithium niobate \cite{k_suzuki_high-speed_2007,he_high-performance_2019}. Fig.\ref{fig:platformcomparison} clearly shows that given an arbitrary loss, SiN is the most scalable platform allowing for the largest useful processor size.

Silicon nitride is also a highly versatile platform enabling on-chip quantum light sources \cite{uppu_-chip_2020,zhao_microresonator_2020, gyger_reconfigurable_2021}, fast switching \cite{epping_ultra-low-power_2017}, lowest propagation losses \cite{bauters_planar_2011} and superconducting single-photon detectors \cite{schuck_nbtin_2013,schuck_quantum_2016}. Together with the highest capability of large-scale linear optical interferometer, as shown in this work, silicon nitride presents itself as a viable path to fully integrated quantum computers.

In conclusion, we have demonstrated a record universal quantum photonic processor with 20 input/output modes, which is the largest processor to date. Thanks to its low loss and high fidelity operations, it is the pinnacle of all universal quantum photonic processors demonstrated thus far.

\begin{acknowledgments}
Funding is acknowledged from the Nederlandse Organisatie voor Wetenschappelijk Onderzoek (NWO) via QuantERA QUOMPLEX (Grant No. 680.91.037), NWA (Grant No.40017607), and Veni (grant No. 016.Veni.192.121).
\end{acknowledgments}

\appendix
\section{}
We report the measurements of the insertion loss for each mode of the processor. Light is injected at each $input_i$ and detected at its corresponding $output_i$. To achieve this, each unit cell of the processor is set to have output light in the same waveguide as the input.
\begin{figure}[h!tb]
  \includegraphics[scale=0.94]{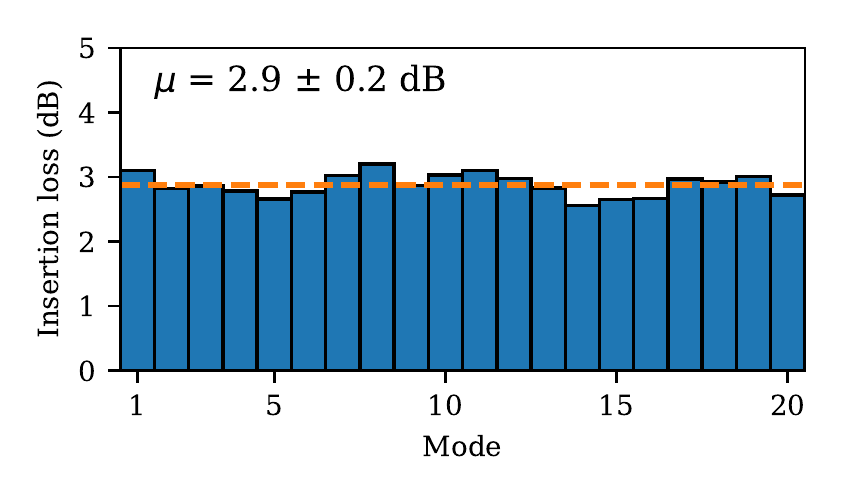}
  \caption{Bar plot of the insertion loss per mode number.}
  \label{fig:losses}
\end{figure}
\section{}
A Gaussian fit of a single HOM dip is shown where a baseline was establish by taking the average of all points that were more than two standard deviations away from the center of the dip. 
\begin{figure}[h!tb]
  \includegraphics[scale=0.94]{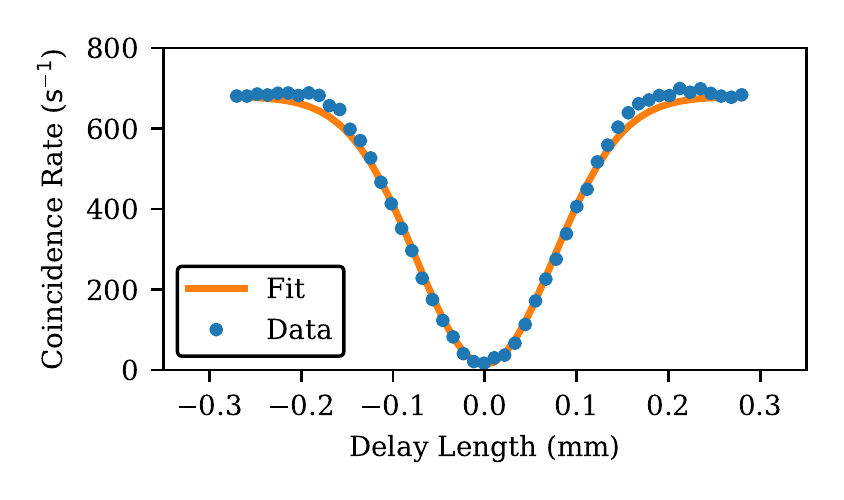}
  \caption{Example of a Gaussian fit for one of the measured HOM dips.}
  \label{fig:hom-fit}
\end{figure}
\clearpage
\bibliographystyle{quantum}
\bibliography{references2}

\end{document}